\documentstyle[aps]{revtex}

\begin{document}
\title{Does General Relativity Require a Metric}
\author{James L. Anderson}
\address{Stevens Institute of Technology\\
Hoboken, New Jersey 07666, USA}
\maketitle

\begin{abstract}
The nexus between the gravitational field and the spece-time metric was an
essential element in Einstein's development of General Relativity and led
him to his discovery of the field equations for the gravitational
field/metric. We will argue here that the metric is in fact an inessential
element of this theory and can be dispensed with entirely. Its sole function
in the theory was to describe the space-time measurements made by ideal
clocks and rods. However, the behavior of model clocks and measuring rods
can be derived directly from the field equations of general relativity using
the Einstein-Infeld-Hoffmann (EIH) approximation procedure. Therefore one
does not need to introduce these ideal clocks and rods and hence has no need
of a metric.
\end{abstract}

\pacs{abc}

\section{INTRODUCTION}

Both the Newtonian and special re4lativistic descriptions of dynamical
systems require the introduction of absolute objects for their formulation.
Newtonian laws use planes of absolute simultaneity and straight lines while
special relativistic laws employ light cones and straight lines. These
elements are absolute in the sense that they are unaffected by the presence
or behavior of any kind of physical system.\cite{PRP} They also characterize
the geometry of Newtonian and special relativistic space-time. In the case
of special relativity, a specification of the light cone structure and the
time-like straight lines is equivalent to the introduction of a flat
Lorentzian metric onto the four dimensional space-time manifold. And
finally, these objects are the source of all inertial effects such as
Coriolis and centripetal forces.

Einstein's famous elevator gedenken experiment led him to his Equivalence
Principle which asserts the local equivalence of inertial and gravitational
effects. This equivalence led him in turn to associate the gravitational
field with the space-time metric. His great achievement was to realize that,
since the gravitational field is a dynamical object, the metric must
therefore also be dynamical. His search for field equations for this
gravitational field/metric took a number of twists and turns but in 1915 he
published the details of his General Theory of Relativity which included his
now well-known field equations. Today these equations are almost universally
accepted and in the last few years their experimental testing has left
little doubt concerning their validity in the macroscopic realm. Since
Einstein first introduced his theory there have been numerous attempts at
alternate theories of gravity, none of which have survived. At the same
time, the success of Einstein's ''geometrization'' of gravity led him and
many other physicists including Weyl, Schr\"{o}dinger, Kalutza and Klein to
attempt a geometrical unification of gravity and electromagnetism. Most of
these attempts survive now as relics of a bygone, simpler age when there
were only two known fundamental forces.

As originally conceived, general relativity consisted of several disjoin
parts. There were the field equations for the gravitational field/metric
tensor $g_{\mu \nu }$ together with prescriptions on how to couple this
field to other fields, e.g., the electromagnetic field. In addition it was
assumed that otherwise free particles followed time-like geodesics
determined by the $g_{\mu \nu }$ while light rays followed null geodesics.
And finally the metric character of $g_{\mu \nu }$ was used to determine the
results of space-time measurements made with so-called ideal rods and clocks
through the introduction of the line element 
\begin{equation}
ds^2=g_{\mu \nu }dx^\mu dx^\nu .  \label{1.1}
\end{equation}
(Here and in what follows Greek indices take the values 0,1,2,3, Latin
indices take the values 1,2,3 and I use the Einstein summation convention.)
In particular, ideal clocks were assumed to measure the integral of $ds$
along their world line.

At this point several comments are in order. First, the Einstein field
equations do not depend on a metric interpretation of $g_{\mu \nu }$: if one
requires that these equations be linear in second derivatives of the
components of $g_{\mu \nu }$, that they contain no absolute objects, are
generally covariant and follow from a variational principle then they are
unique modulo the so-called cosmological term. Second, the geodesic
equations, even when applied to test masses, can only be approximate since
any finite mass will in general radiate in the course of its motion through
a gravitational field. There is also the question of what field to use to
calculate the geodesics. The whole procedure only makes sense if one uses
the background gravitational field/metric in which the mass moves. But
because the field equations are highly non-linear such a separation between
an external field and the field of the mass can only be accomplished by the
use of some approximation procedure. Furthermore, the geodesic hypothesis
can say nothing about the motion of two comparable masses. And finally,
there is a problem with the proper time hypothesis - how does one identify
an ideal clock. Any real physical clock will be subject to tidal forces
which, if they are sufficiently strong, can disrupt or at least perturb its
workings and hence will cease to be an ideal clock. Thus any real clock is
at best an approximation to an ideal clock.

These of course are not new problems - they have been around as long as
general relativity itself. However, what is not well known is that in 1949
Einstein, together with Infeld and Hoffman \cite{EIH} and later with Infeld%
\cite{EI} (collectively EIH) laid the groundwork for their resolution. For
what these authors showed was that the motion of compact sources of the
gravitation field followed from the field equations themselves without the
need for additional assumptions such as the form of a force law such as one
must make in electrodynamics. Even the motion of electrically charged
particles was shown to follow from the combined Einstein-Maxwell field
equation.\cite{JLA1} without the need to postulate the Lorentz force law. In
particular it is unnecessary to make the geodesic hypothesis since an
approximate form of the geodesic equations follows directly from the field
equations. Furthermore, it is possible to construct simple clock models,
essentially two compact masses or charges moving around each other in
circular orbits, whose dynamics in an external field can again be determined
directly from the field equations.\cite{JLA2} As a consequence one does not
need to use the metric interpretation of $g_{\mu \nu }$ for any purpose and
hence the notion of a metric can be dispensed with entirely in general
relativity.\cite{JLA4}

While this view may appear to be quite radical I would characterize it
rather as conservative since it sets forth the minimum number of elements
needed to formulate general relativity. Furthermore, it is important in
understanding a theory, particularly a fundamental theory, to know what is
essential to that theory and what is not. This is not to say that the
geometrical interpretation is of no use as it is often of heuristic value in
affording a picture of what is going on in a particular situation. And of
course many of the concepts of differential geometry such as Killing
vectors, Lie derivatives etc. are extremely useful in the analysis of
particular gravitational fields. And finally, it does not detract in any way
from Einstein's great achievement in constructing the general theory of
relativity.

In what follows I will outline briefly the main ideas behind the EIH
procedure and how one can use it to determine the behavior of simple clock
models in an external gravitational field. Since most of this material is
already in the literature I will try to avoid unnecessary details as much as
possible.

\section{EQUATIONS OF MOTION}

Perhaps Einstein's least well known achievement, but arguably one of his
most important, is the work he did with L. Infeld and B. Hoffmann on the
equations of motion of sources of the gravitational field. General
relativity is unique among field theories in that one does not have to
postulate separately the equations of motion of the sources of the fields
they produce. In classical electrodynamics one must postulate not only the
Lorentz force but the equations of motion in which it appears as well. In
general relativity neither of these types of postulates are necessary.
Furthermore, in deriving these equations one does not encounter self energy
infinities found in other derivations.

The equations of motion can most easily be derived from a form of the field
equations given by Landau and Lifshitz \cite{LL} which are 
\begin{equation}
U^{\mu \nu \rho }{}_{,\rho }=\Theta ^{\mu \nu }\text{ ,}  \label{2.1}
\end{equation}
where $U^{\mu \nu \rho }$ is a so-called superpotential, a function of $%
g_{\mu \nu }$ and linear in its first derivatives and 
\begin{equation}
\Theta ^{\mu \nu }=(-g)(T^{\mu \nu }+t_{LL}^{\mu \nu }).  \label{2.2}
\end{equation}
In this latter equation $g=$det$(g_{\mu \nu })$, $t_{LL}^{\mu \nu }$ is the
Landau-Lifshitz energy-stress pseudotensor and $T^{\mu \nu }$ is
energy-stress tensor due to the presence of other fields. A contribution to
this latter tensor due to the sources of these and the gravitational field
is not included because these sources are assumed to be compact and vanish
on the closed surface integrals used in the EIH derivation. Because of the
antisymmetry of $U^{\mu \nu \rho }$ in $\nu $ and $\rho $, $U^{\mu
rs}{}_{,s} $ is a curl whose integral over any closed spatial 2-surface
vanishes identically. As a consequence, integration of Eq. (2.1) over such a
surface in a $t=$ constant hypersurface gives 
\begin{equation}
\displaystyle \oint %
\,(U^{\mu r0}{}_{,0}-\Theta ^{\mu r})n_r\,dS=0,  \label{2.3}
\end{equation}
where $n_r$ is a unit surface normal. It is this last equation that yields
the equations of motion of a source when the surface encloses it.

The actual details of deriving these equations requires the use of a number
of approximation procedures and the identification of the small parameters
associated with the system under consideration.\cite{JLA3} For the types of
systems to which one can apply the EIH method, one of these parameters, $%
\epsilon _S$, is a characteristic time scale $T_S$ of the system such as its
period, divided by the light travel time $T_L$ across the system and is
equivalent to a characteristic velocity of the system divided by the speed
of light. Since this ratio must be small compared to one, the approximation
is referred to as a slow-motion approximation. I will not attempt here to
give the details of the derivation since they are rather lengthy and refer
the interested reader to the references cited below. If one perturbs off a
''flat'' gravitational field ($g_{\mu \nu }=$ diag$\,(1,-1,-1,-1)$ ) as was
done in the original EIH papers one obtains, in first order or Newtonian
approximation, the equations of motion, including the inertial force term,
of compact sources interacting via the $1/r^2$ Newtoian force law. Since
there is only one mass parameter per source in these equations one also
derives the equivalence of inertial and gravitational mass by these means.

In ref.\cite{JLA2} I the EIH method to derive the approximate laws of motion
of compact masses, both charged and uncharged, in a Einstein-deSitter
gravitational field given by 
\begin{equation}
g_{\mu \nu }=\text{ diag\thinspace }[1,-R^2(t),-R^2(t),-R^2(t)]  \label{2.4}
\end{equation}
where $R(t)=(t/t_0)^{2/3}$ and $t$ is the cosmic time. In this case a second
small parameter $\epsilon _H\ll \epsilon _S$, the ratio of $T_S$ divided by
the Hubble time $T_H=$ $R(t)/\dot{R}(t)$ is used in a multi-time
approximation scheme. In this scheme the source coordinates are assumed to
depend both the the cosmic time $t$ and a fast time $\tau =t/\epsilon $
where $\epsilon =\epsilon _H/\epsilon _S$. The resulting equations of motion
in lowest order of approximation in $\epsilon _S$ and $\epsilon _H$ take the
form
\begin{equation}
m_A{\bf x}_{A\tau \tau }+2\,\epsilon \,m_A{\bf x}_{\tau t}+2\,\epsilon \,m_A%
\frac{R_t}R{\bf x}_A=-\frac 1{R^3}\sum_{B\neq A}\frac{m_Am_B}{r_{AB}^3}\,%
{\bf r}_{AB}  \label{2.5}
\end{equation}
where ${\bf r}_{AB}={\bf x}_A-{\bf x}_B$. For a two-body system moving in
circular orbits about each other one finds that  
\begin{equation}
\omega =\text{constant \hspace{0pt}\qquad and\qquad }rR(t)=\text{constant }
\label{2.6}
\end{equation}
where $\omega $ is the angular frequency of the motion and $r$ is the
coordinate radius of the orbit. We see that such a system can serve as a
clock and such a clock measures the cosmic time $t$. This result is valid
both for charged and uncharged clocks so, in the level of approximation,
there is no difference in the time measured by a ''gravitational'' and an
''electrical'' clock. However, this will not be true in higher orders of
approximation. Unfortunately, the next corrections in $\epsilon _H$ are of
order $\epsilon _H^2$ and so are probably not observable at the present
time. Note that, as $R(t)$ increases in time, the coordinate radius of a
clock decreases. Depending on whether we choose to take the coordinate
radius or the coordinate radius times the scale factor to be a measure of
the size of the clock we can say either that the size of the clocks, viewed
now as a measuring rod, is decreasing while the size of the universe remains
fixed or else that the size of the clocks stays fixed while the size of the
universe increases. There is no way to distinguish physically between the
two interpretations. With either interpretation however the number of such
rods needed to measure the distance to a distant galaxy will increase in
time. We also see that small systems such as the solar system or atoms with
short time scales compared to the Hubble time will effectively not see the
effects of expansion while big ones for with $\epsilon _H\sim 1$ will.

\section{CONCLUSIONS}

In this paper I have argued that a metric interpretation is not needed in
general relativity and that the purposes for which it was originally
introduced, i.e., temporal and spatial measurements and the determination of
geodesic paths, can be all be derived from the field equations of this
theory by means of the EIH approximation scheme. As a consequence, the only 
{\it ab initio} space-time concept that is required is that of the blank
space-time manifold. In this view what general relativity really succeeded
in doing was to eliminate geometry from physics. The gravitational field is,
again in this view, just another field on the space-time manifold. It is
however a very special field since it is needed in order to formulate the
field equations for, what other fields are present and hence couples
universally with all other fields. It is hoped that this identification of
unnecessary concepts in general relativity will help in the discussion of a
number of fundamental issues, chief among which is that of construction a
quantum theory of gravity.

\end{document}